\documentclass[prl,twocolumn,amsmath,amssymb,floatfix,showpacs]{revtex4}
\usepackage{graphicx}
\usepackage{bm}
\usepackage{amssymb}
\usepackage{color}
\usepackage{epsfig}
\usepackage{subfigure}
\usepackage{draftcopy}

\newcommand{\be}{\begin{equation}}
\newcommand{\ee}{\end{equation}}
\begin{document}

\title {Localization, anomalous diffusion and slow relaxations: a random distance matrix approach}

\author{Ariel Amir, Yuval Oreg, Yoseph Imry}

\affiliation { Department of Condensed Matter Physics, Weizmann
Institute of Science, Rehovot, 76100, Israel\\}
\begin {abstract}
We study the spectral properties of a class of random matrices where
the matrix elements depend exponentially on the distance between
uniformly and randomly distributed points. This model arises
naturally in various physical contexts, such as the diffusion of
particles, slow relaxations in glasses, and scalar phonon
localization. Using a combination of a renormalization group
procedure and a direct moment calculation, we find the eigenvalue
distribution density (i.e., the spectrum), for low densities, and
the localization properties of the eigenmodes, for arbitrary
dimension. Finally, we discuss the physical implications of the
results.

\end {abstract}

 \maketitle

Application of the theory of random matrices whose elements are
independent Gaussian variables has proven to be rich mathematically
and relevant for many physical systems \cite{mehta}. In this Letter
we study a different class of random matrices where the $i, j$'th
element is a function of the Euclidian distance $r_{ij}$ between
pairs of points whose positions are chosen randomly and uniformly in
a $d$-dimensional space. It is natural that in cases where the
matrix element is related to an overlap between localized
quantum-mechanical wavefunctions, the dependence on the distance
will be exponential, i.e., $A_{ij}=e^{-r_{ij}/\xi}$, with $\xi$
being the localization length \cite{lifshitz}.

The exponential matrix is an appropriate model for various physical
systems, in this Letter we will concentrate on its application  to
glasses relaxing to equilibrium, a particle diffusing in random
environment and localization of phonons. Most of the results are
derived at the low density limit, when $\epsilon=\xi/{r_{nn}} \ll
1$, with $r_{nn}$ being the average nearest neighbor distance. To
understand the properties of these systems one need to find out the
distribution density $P(\lambda)$ of the eigenvalues ${\lambda}$ as
well as the structure of the eigenmodes. An intuitive picture of the
problem arises in the application to phonon localization with
springs constants $K_{ij}$ that depend exponentially on the
Euclidean distances between the masses; we therefore use the phonon
terminology: eigenmode.

\begin{figure}[b!]
\includegraphics[width=0.5\textwidth]{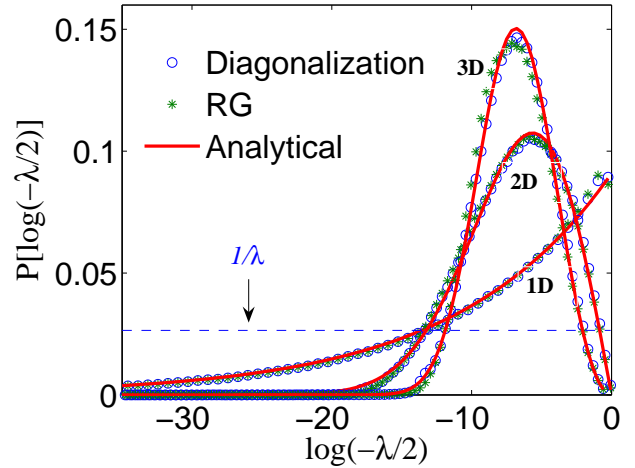}
\caption{Comparison between the exact numerical diagonalization
(circles), and different theoretical approaches: The stars (green)
show the results of a renormalization group approach, where, if one
considers the problem as finding the normal modes of a spring
network, the springs and masses are renormalized. The solid line
(red) depicts the analytical results of Eq. (\ref{prob_general}), and Eq. (11) of the SM.
The numerical results shown are for 1, 2 and 3 dimensions, with
N=1000, averaged over 1000 realizations. The points were chosen in a
line, square or box of side one, and $\epsilon=0.1$, corresponding
to $\xi=10^{-4}$ in 1D, $\xi=0.0032$ in 2D and $\xi=0.01$ in 3D.
Notice that the graph shows the distribution density of the
\textit{logarithm} of the eigenvalue, which eliminates the governing
$1/\lambda$ dependence, and allows us to observe clearly the
deviations from it. To demonstrate this, the horizontal dashed line
(blue) shows an exact $1/\lambda$ distribution. No fitting
parameters are used. \label{rg_2d} }
\end{figure}

The low density limit allows us to find $P(\lambda)$ analytically
employing a direct calculation of its moments, see Eq.
(\ref{prob_general}) and the Supplementary Material (SM). We find
that $P(\lambda) \sim 1/\lambda$ in all dimensions over a broad
range of $\lambda$'s. While in one dimension the normalization of
$P(\lambda)$ is assured by an integrable power-law divergence at
eigenvalues close to zero, for higher dimensions there is a peak
related to a finite cutoff, cf Fig. \ref{rg_2d}. We use a
logarithmic scale to plot $P[\log (-\lambda/2)]$ in order emphasize
the deviations from the $1/\lambda$ distribution.

\begin{figure}[b!]
\includegraphics[width=0.55\textwidth]{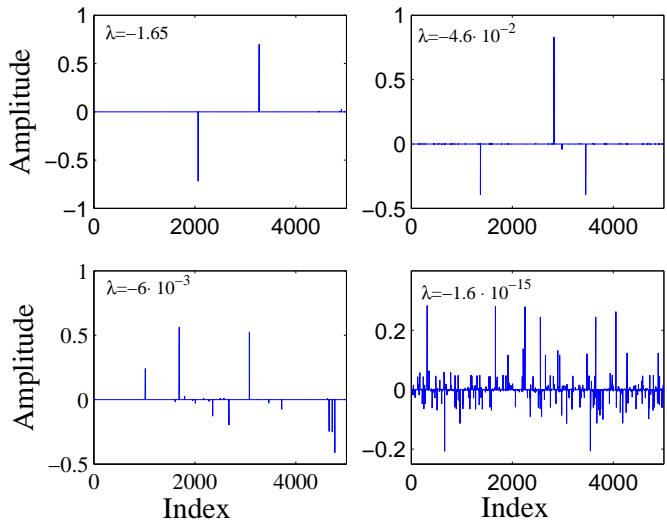}
\caption{Demonstration of the structure of the
eigenmodes. $N=5000$ points were chosen randomly and uniformly in a
two dimensional box, with $\epsilon= \xi/r_{nn} =
0.1$. The eigenmodes become more delocalized as the eigenvalues
approach zero, a condition made quantitative in Eq. (\ref{nc}), through the use of the RG approach.
Eigenmodes are comprised of clusters of points, localized in real
space. \label{loc} }
\end{figure}

To comprehend the structure of the eigenmodes we use a
renormalization group (RG) approach for random systems that was
developed in the context of spin chains
\cite{dasgupta,fisher,altman,gil}. At each RG step, we choose the
stiffest spring. Since the spring is large by construction, after
finding the eigenvalue associated with the stiffest spring we can
'glue' together the two masses at its ends creating a larger mass.
At the next RG step we choose again the stiffest spring among those
who remain. In this way the large eigenmodes are built initially by
pair of masses, but as the RG process progress larger clusters of
masses form eigenmodes with smaller eigenvalues. This behavior is
demonstrated in Fig. \ref{loc}.

Since in the RG step a spring is chosen regardless of the masses,
the flow of the mass distribution is independent of that of the
springs. Namely, we will get statistically the same distribution of
clusters as a function of the RG step by a coagulation process where
we randomly choose two masses and glue them together at each step.
This observation allows us to find the mass distribution as a
function of the number of RG steps (see Eqs. (7) and (8) of the SM).
Together with the exact result for the eigenvalue distribution, we can find the dependence of the
cluster mass on the eigenvalue $n_c(\lambda)$, cf Eq. (\ref{nc}).

We will now present in more details the properties of the
exponential random matrix model, its applications, and the
derivation of $P(\lambda)$ and $n_c(\lambda)$

\emph{Model and relevant physical problems.-} $N$ points are chosen randomly in a $d$ dimensional
cube, and $r_{ij}$ is defined as the Euclidean distance between
points $i$ and $j$. We define a matrix $A$, as follows:
$A_{ij}=e^{-r_{ij}/\xi}$, for $i \neq j$ and $A_{ii}=-\sum_{j \neq
i} {A_{ij}}$, the latter definition expressing a conservation law in
the physical problem \cite{distance_matrices,amir_glass, carsten}.
We shall be interested in determining $P(\lambda)$, the probability
density of eigenvalues of the matrix $A$, for low densities (small
values of $\epsilon$). Since the matrix is hermitian, it is clear
that all its eigenvalues are real, and it can also be proven that
they are negative \cite{bogomolny2,amir_glass}.

This model is relevant for problems from various fields of physics,
but for now, we choose to focus on scalar phonon localization. It
will be useful to have this problem in mind when we discuss the RG
calculation.

When studying normal modes of a collection of equal masses $m$
connected by harmonic springs, one has to find the eigenmodes of a
matrix $A$, where $\{A_{ij}\}$ are the spring constants, and due to
momentum conservation the sum of columns vanishes. The eigenvalues
are related to the frequencies by $m \omega^2 = -\lambda$, where we
can choose $m$ to be unity, for convenience. The above model can be
used to study phonons in a disordered lattice, where due to the
randomness in the matrix elements (i.e., a distribution of spring
constants), the oscillating modes may turn out to be localized in
space~ \cite{nagel}. Notice that this is a 'scalar' phonon model
\cite{schirmacher}, where the vectorial properties of the modes have
not been taken into consideration (and thus the matrix has $N$ and
not $N d$ modes).

For phonons, the density-of-states of the vibrations $\tilde
P(\omega)=2 m \omega P(\lambda=-m\omega^2)$ determines the thermal
properties of the system \cite{Vitelli}. It is interesting to note
that our result for $P(\lambda)$ implies $P(\omega) \sim 1/\omega$
up to corrections depending on the dimensionality.

\emph {Derivation of the eigenvalue distribution.-} The exponential
matrices model in one dimension has peculiar spectral properties
\cite{alexander, ziman}. The SM gives a simple, non-rigorous
derivation for the low density limit of the eigenvalue distribution
in one-dimension, which turns out to be a power-law. The argument
also shows that in one dimension the modes are localized, with
typical size depending on $\lambda$ as a power-law.

We will now develop a general formula for the eigenvalue spectrum at
arbitrary dimension, in the low density limit. We shall see that in
one dimension we obtain a power-law, while in higher dimension we
get logarithmic corrections to a distribution $P(\lambda) \sim
1/\lambda$ .

Using the following sum:

\be  I_k \equiv \int P(\lambda) \lambda^k d\lambda = \frac{1}{N}
\left \langle \sum_{i_1,i_2..i_{k}}
A_{i_1,i_2}A_{i_2,i_3}...A_{i_{k},i_1} \right\rangle ,\ee one can calculate the $k$'th moment of the probability density. In the SM we find that in the low density approximation $\epsilon \rightarrow 0$, $I_k \approx -(-2)^{k-1} d! V_d (\epsilon/k)^d,$ with
$V_d=\pi^{d/2}/\Gamma(d/2+1)$ the volume of a \emph{d} dimensional
unit sphere.

Since a distribution is fully determined by its moments under
certain conditions which are fulfilled in our case \cite{moments},
it suffices to find a distribution which yields these moments. It
can be directly checked by performing the integrals that the
following probability density does this:

\be P(\lambda)=  \frac{\epsilon^d d V_d /2
[-\rm{log}(-\lambda/2)]^{d-1} e^{-\frac{V_d}{2} \epsilon^d
(-\rm{log}(-\lambda/2))^d}}{\lambda} . \label{prob_general}\ee

The cumulative $C(\lambda)$ of this distribution takes a
particularly simple form: \be C(\lambda)\equiv \int_\lambda^0
P(\lambda)d\lambda= e^{-\frac{V_d}{2} \epsilon^d
(-\rm{log}(-\lambda/2))^d} \label{generald}. \ee


In one-dimension we find a power-law divergence in the distribution
density, while in a dimension $d>1$ we see from Eq. (\ref{prob_general})
that there is a finite cutoff at small eigenvalues. Fig.
\ref{rg_2d} compares this formula with numerical diagonalization,
for the one, two and three dimensional cases, as well as the RG
procedure.

It is important to distinguish the above result from, for example,
those of Ref. [\onlinecite{amir_glass}], which relates the
eigenmodes to isolated pairs of points. There, an uncontrolled
approximation was used, connecting pairs of nearest-neighbors (a
procedure which is not always well-defined). While giving the
correct qualitative dependence, it differs from Eq. (\ref
{prob_general}) by a factor of 2 in the exponential, as well as in
the normalization factor. As will be shown via the renormalization
group approach in the next section, this difference reflects
important differences in the underlying physics, and in the
structure of the eigenmodes: the high frequency eigenmodes indeed consist of
pairs of nearest-neighbors masses, and for them the two approaches
coincide. However, the low-lying modes are comprised of an
increasing number of masses, diverging as the frequency approaches
zero. For any finite value of $\epsilon$, Eq. (\ref{prob_general})  might be violated for sufficiently small
$\lambda$, while keeping the corrections to all moments of the distribution sufficiently small. 

\emph{Renormalization group approach.-} Following
\cite{dasgupta,fisher,altman,gil}, let us consider the
renormalization group (RG) approach to this problem. This will
enable us to understand the localization properties of the
eigenvectors. We recall the scalar phonon localization picture,
where the points represent masses and the matrix-elements represent
spring constants. At \emph{each} RG step, we choose the largest
spring $K$: due to the broadness of the spring distribution, it will
not be affected much by the neighboring springs, and therefore it
will contribute a frequency $\omega^2 =\frac{K}{\mu}=-\lambda$,
where $\mu=\frac{m_1 \cdot m_2}{m_1+m_2}$ is the reduced mass of the
two masses it connects, $m_1$ and $m_2$. Initially, all masses are
equal. Notice that the mechanical intuition tells us we should
choose the largest \emph{spring} at each step, and not necessarily
the largest \emph{frequency}, i.e., the choice of the springs does
not depend on the masses. Since this spring is large by
construction, we can now 'glue' the two masses together, and create
a single mass $m_1+m_2$. We also have to renormalize the springs
attaching the new mass and all other masses: a reasonable way to do
so is to take each of the springs between a mass $m$ and the new
(joint mass) as the sum of the two springs between the mass $m$ and
each of the masses $m_1$ and $m_2$. Clearly we will obtain smaller
frequencies and springs and larger masses as the RG progresses.

 As is shown in Fig. (\ref{rg_2d}), this simple scheme captures the
essential physics, with no fitting parameters. As
mentioned, the reason the method works so well is the broadness of
the 'spring' distribution: for a one dimensional case, for example,
the distribution of the nearest-neighbor spring constants (which can
be calculated directly from the exponential distribution of the
distance intervals) follows a $P(K) \sim 1/K^{1-\epsilon}$, where
$\epsilon \rightarrow 0 $ in the low density limit. Notice that for
the one-dimensional case the RG procedure would choose exactly these
nearest-neighbor springs, by construction. Thus, were we to neglect
the mass RG we would obtain $P(\lambda) \sim
1/\lambda^{1-\epsilon}$, which recovers the low density result
mentioned earlier. In the SM we show how one can incorporate the
mass RG to correct the one-dimensional result also for higher
densities.

 Since at each RG step
a spring is chosen regardless of the masses, the flow of the mass
distribution is completely independent of that of the springs. As mentioned, the results of the mass RG are applicable
for any dimension. We will now show that using the probability
density we obtained, Eq. (\ref{prob_general}), with the results of
the mass RG, we can understand the localization properties of the
eigenvectors. The SM shows how we can find the distribution of masses at a given
stage of the RG process, which turns out to be approximately an
exponential distribution $e^{-m/\langle m \rangle}$, with $\langle m
\rangle$ changing as the RG process evolves, corresponding to the
formation of larger and larger clusters. This implies there is a typical mass $\langle m \rangle$ at each instance. Since at each step of the
RG process the number of clusters decreases by one, after $k$ steps
the average cluster mass is given by: $\langle m \rangle = N/(N-k)$,
which is also the typical size of the cluster, $n_c$. On the other
hand we can find the relation between the RG step $k$ and the
eigenvalue $\lambda$: In the RG process at each step the number of
masses is decreased by one, and the corresponding eigenmode
recorded, starting with the highest springs (and eigenvalues). At
the stage of the RG flow corresponding to an eigenvalue $\lambda$,
the number of masses left, $N-k$, can be calculated using Eq.
(\ref{generald}):

\be k=  N\int_{-2}^{\lambda} P(\lambda) d\lambda =N(1-C(\lambda)).
\label{combine} \ee

Combining this equation with that for $n_c=\langle m \rangle$, we
find that $n_c$ depends on the eigenvalue as:

\be n_c(\lambda) \sim 1/C(\lambda) =e^{\frac{V_d}{2} \epsilon^d
(-\rm{log}(-\lambda/2))^d}. \label{nc} \ee As we go to zero
eigenvalue, the size of the eigenmodes diverges, as demonstrated in
Fig. (\ref{loc}) on a particular example. Notice that for the case
$d=1$ we recover the power-law relation between localization length
and $\lambda$ mentioned earlier, and related to Refs.
[\onlinecite{alexander}] and [\onlinecite{ziman}].

Eq. (\ref{combine}) also allows us to count $\mathcal{N}(n_c)$, the
number of all clusters containing more than $n_c$ masses: their
number is $N \int_\lambda^0 P(\lambda)d\lambda= N C(\lambda)$, where
we know the dependence of $\lambda$ on $n_c$ through Eq. (\ref{nc}).
This gives: $\mathcal{N}(n_c)/N = 1/n_c$.

%
%

\emph{Physical implications.-} The mathematical model presented is
relevant also for other physical problems besides phonon
localization discussed above. For example, one can consider the
hopping of a particle in a random environment, where $A_{ij}$
describes the transition probability of a particle from site $i$ to
site $j$. If we define the probabilities of the particle to be at
the different sites by a vector $\vec{p}$, then
$\frac{d\vec{p}}{dt}=A \vec{p}$. In disordered systems, the
transition rate often depends exponentially on the distance
\cite{scher}. Another physical example of relevance, is the study of
relaxations in glasses. Under certain approximations, this can be
mapped to the study of eigenmodes of a class of random matrices
related to the one described above \cite{amir_glass_related} and in
fact this has been the original motivation for this study. In both
the cases of the diffusion problem and the relaxations in glasses
the Laplace transform of the distribution density plays an important
role: in the hopping problem, it gives the probability to remain in
the origin \cite{diffusion_laplace}, while for the glass relaxation
it corresponds to the time dependence of the relaxation
\cite{amir_glass, pollak2}.

Upon taking the Laplace transform $\hat{P}(t)$ of the distribution
density in one-dimension (with argument $t$), we obtain $\hat{P}(t)
\sim t^{-\frac{\epsilon}{1+\epsilon}}.$ This implies that the
diffusion in this case is anomalous \cite{{scher},{anomalous1},
{anomalous2}}: the probability to remain in the origin does not
decrease as $1/\sqrt{t}$, as is the case for normal diffusion, but
as a smaller power (the particle tends to be more localized). If we
assume that after a time $t$ the particle has spread over a distance
$r$, it is reasonable to assume that $\hat{p}(t)r \sim 1$, showing
that $r \sim t^{\frac{\epsilon}{1+\epsilon}}$. In the case $\epsilon
\rightarrow 0$, this can be approximated as $ \hat{P}(t) \sim
C-\rm{log}(t)$. This type of behavior has been experimentally
observed in various types of glassy systems \cite
{{ludwig2},{Grenet1},{zvi:3}}. In higher dimensions, one obtains in
the low density limit $\hat{P}(t)=\phi[\rm{log}(t)]$, with $\phi$ a
polynomial of degree $d$. This will be elaborated on in future
works.

\emph{Summary.-} We presented here a model of random matrices which
captures the interesting physics of various different systems. After
introducing the model, we found the eigenvalue distribution density
in the low-density limit (Eq.~(\ref{prob_general})) and the
localization properties of the eigenmodes (Eq.~(\ref{nc})) using a
direct moment calculation as well as a renormalization group
approach. Our results for the spectrum agree with the 1D case \cite{alexander,ziman}, and with the exact
numerical diagonalization. While in one dimension it is known that
for an infinite system there will be an (integrable) power-law
divergence of the spectrum, in a higher dimension $d$ we found that
there is a finite cutoff, where
$\epsilon=\xi/r_{nn}$ is the small parameter of the
theory. We used the RG approach to show that the eigenmodes are
localized, and to find a relation between the spatial extent of an
eigenmode and the corresponding eigenvalue, implying that this size
diverges as the eigenvalue approaches zero.

We discussed the
application of the model for various physical problems, such as
relaxations in glasses, diffusion of particles in random media, and
localization. In the future, it would be fascinating to understand
also the crossover or phase-transition to the high density or low
disorder regime, which should present different physics.

We thank M. Aizenman, D. Cohen, M. E. Fisher, O. Hirschberg, B.
Nedler, W. Schirmacher, V. Vitelli and O. Zeitouni for important
discussions. This work was supported by a BMBF DIP grant as well as
by ISF and BSF grants and the Center of Excellence Program.

\section{Supplementary Material}

\maketitle
\section {1. A nonrigorous derivation in one-dimension}

It is known that for an isolated system the center-of-mass motion
gives a zero eigenvalue mode, and thus in our model a cluster of
points largely separated from the rest of the points will give an
exponentially vanishing eigenvalue. Let us consider the statistics
for finding such a cluster, isolated on both sides by a distance $r$
(here we explicitly use the one dimensional character of the
problem).

By perturbation theory the vanishing eigenvalue will be shifted by a
value of order $|\lambda| \sim e^{-r/\xi}$, where $\xi$ was the
decay length of the matrix elements. We choose the points from a
uniform distribution, hence the distances between them are
distributed exponentially. Therefore, if we move (e.g. to the right)
along the one dimensional chain, the probability of encountering a
gap of size $\tilde r$ larger than $r$ is $P(\tilde r>r) =e^
{-r/r_{nn}}$ with $r_{nn}$ being the average distance between the
points. If we proceed to move along the chain we are certain to
encounter a gap larger than $r$ eventually. Thus, the probability
density of finding a cluster isolated from both sides by a distance
$r$ (obtained by differentiating the cumulative distribution) is
$p(r) =e^{-r/r_{nn}}/r_{nn} $. Using the relations, $P (\lambda) d
\lambda = p(r) dr$ we find $P(\lambda) \propto 1/\lambda^
{1-\epsilon}$, with $\epsilon=\xi/r_{nn}$ being a small parameter if
the density of points is low enough. This analysis also allows us to
find the length of the cluster, since the typical number of points
we have to pass until we encounter the second gap is $l(\lambda)=1/p
\propto e^{r/r_{nn}}\sim \lambda^{-1/\epsilon}$. This reproduces the
low density limit of a similar problem \cite{alexander,ziman},
albeit with nearest-neighbor matrix elements only (indeed,in the low density limit, in one dimension, we do not expect
next-nearest neighbor couplings to modify the results).
 The relation $l(\lambda)$ that was obtained implies
that the spatial extent of the modes diverges as the eigenvalue
approaches zero.

\section {2. Calculation of the moments of the eigenvalue probability
density}

We would like to calculate the $k$'th moment of the probability
density using the following sum:

\be \int P(\lambda) \lambda^k d\lambda = \frac{1}{N}\langle
\sum_{i_1,i_2..i_{k}} A_{i_1,i_2}A_{i_2,i_3}...A_{i_{k},i_1} \rangle
\ee
To gain intuition, let us consider the first moment: $ \int
P(\lambda) \lambda d\lambda = \frac{1}{N}\langle \sum_{i} A_{i,i}
\rangle. $ By construction, the magnitude of the diagonal elements
is the sum of all the other elements in the columns, i.e., sum of
the hopping matrix elements to all neighbors, close and far. To the
lowest order in $\epsilon$, we can consider only the
nearest-neighbor of each point, as we shall shortly demonstrate.
Under this approximation the first moment can be readily calculated:
$ I_1 \equiv \int P(\lambda) \lambda d\lambda \approx -\int_0^\infty
P_{nn}(r)e^{-r/\xi} dr, $where $P_{nn}(r)$ is the nearest-neighbor
distance distribution, which is given by $P_{nn}(r)=\frac{d
V_d}{r_{nn}} {(r/r_{nn})}^{d-1} e^{-V_d {(r/r_{nn})}^d}$, where
$V_d$ is the volume of a $d$ dimensional sphere and $r_{nn}$ is the
average nearest-neighbor distance. In the low density limit this
gives $I_1 \approx -d! V_d \epsilon^d.$ Calculating the corrections
to this formula due to all neighbors, one obtains a correction of
the order of $O(\epsilon^{2d})$, which is indeed of higher order in
the small parameter $\epsilon$. The crux of the matter here is that
due to the exponential function, the contributions to the integral
arise mainly from rare pairs which are very close to each other,
with a distance of order $\xi$ between them. For such pairs the
probability that a third point will be in the vicinity of the two
points is negligible.

\begin{figure}[h!]
\includegraphics[width=0.15\textwidth]{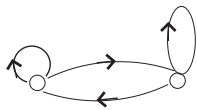}
\caption{An example of a diagram contributing to the $k$'th moment
in the low density limit. \label{diagram2} }
\end{figure}

For the $k$'th moment the general term contributing will be a
combination of hops to the nearest-neighbor and loops that stay in
either the original site or the nearest-neighbor, as illustrated
schematically in Fig. \ref{diagram2}. Thus, there would be $2^{k-1}$
contributing diagrams,  each of equal contribution. Therefore the
$k$'th moment will be given by $I_k \equiv \int P(\lambda) \lambda^k
d\lambda \approx -(-2)^{k-1} \int_0^\infty P_{nn}(r)e^{-k r/\xi}
dr.$ This yields in the low density limit: $ I_k \approx -(-2)^{k-1}
d! V_d (\epsilon/k)^d . \label{momentk}$ These moments correspond to
the probability density given in Eq. (2), as can be verified by a
direct calculation of its moments.

~

 It is interesting to note that although
the leading order contribution to the \emph{k}'th moment was made by
a small subset of configurations, close pairs, the resulting
distribution captures the probability density of eigenvalues whose
corresponding eigenmodes are actually clusters. The clusters can
understood by the renormalization group approach, discussed in the
manuscript. It is generally true that for large \emph{k}'s, the
moments are dominated by the largest (i.e., most negative)
eigenvalues, which in our case are exactly the close pairs (which
are nearly eigenmodes). The above calculation shows that in the low
density limit also the low moments (e.g: $I_1$, above), are
dominated by the close pairs. Thus, we can find all moments and the
complete distribution, by studying the properties of the close
pairs, in a similar fashion to an analytical continuation, by which
studying the function in a small region determines its behavior
further away.

In the case of matrix elements decaying not exponentially but as
some function $f(r)$ of the distance, we can repeat the previous
argumentation. If $f(r)$ decays, for example, faster than
exponential, than surely the contribution of the diagrams neglected
will be smaller still, thus justifying the method. In this case the
\emph{k}'th moment will be: $I_k \equiv \int P(\lambda) \lambda^k
d\lambda \approx -(-2)^{k-1} \int_0^\infty P_{nn}(r)f(r)^k dr.$ It
can be checked that the cumulative of the associated distribution in
this case is given by:

\be C(\lambda)\equiv \int_\lambda^0 P(\lambda)d\lambda=
e^{-\frac{V_d}{2 r_{nn}^d} ((f^{-1}(-\lambda/2))^d},\ee where
$f^{-1}$ is the inverse function of $f(r)$. As an example, applying
this formula for gaussian matrix elements in two-dimensions gives a
power-law distribution density, $P(\lambda) \sim 1/\lambda^{1-\pi
\epsilon^2/2}$.

\section {3. Solution for the renormalization of the masses}

As explained in the manuscript, the RG procedure chooses the largest springs at every instance. With regard to the masses this is a random process not depending explicitly on the masses, as long as we neglect effects arising from the larger surface area of the clusters. As is shown in the manuscript, these do not affect $\langle m \rangle$. We shall now analyze the resulting flow of the mass distribution, under this approximation. Initially, we have a set of unit masses, and at each step we choose two masses at random and combine them. If we denote the \emph{average} number of masses of mass m in the \emph{k}'th stage as $N^k_m$ (at which step we have $N-k$ masses left), we therefore have:

\be N^{k+1}_m \approx N^k_m -\frac{2N^{k}_m}{N-k} + \sum_{m'=1..(m-1)}\frac{N^k_{m'}}{N-k} \frac{N^k_{m-m'}}{N-k} \label {num}\ee

Indeed, the probability that a chosen mass is of mass $m$, is $\frac{N^{k}_m}{N-k}$. This relates to the first term, which accounts for the average reduction in the number of masses of mass $m$ (the factor of two arising from the fact that we choose two masses at each stage). The second term accounts for all the possible ways that two smaller masses can combine to contribute to the number of masses of mass $m$, where again the same form for the probabilities enters.

In order to move from numbers to probabilities, we have to divide by the current number of masses, which is $N-k-1$ (since at each RG step, when two masses are combined to create one mass, we lose one mass). It is useful to write, for $N-k \gg 1$:

\be \frac{1}{N-k-1} \approx \frac{1}{N-k}+\frac{1}{(N-k)^2} . \label {prob}\ee

Denoting by $\mathcal {P}^k_m$ the probability to have a mass of
size $m$ after $k$ springs are eliminated, we have $P^k_m = \frac{N^k_m}{N-k}$, leading to:

\be P^{k+1}_m  \approx P^k_m+\frac{P^k_m}{N-k} -\frac{2P^{k}_m}{N-k} + \frac{1}{N-k}\sum_{m'=1..(m-1)} P^k_{m'} P^k_{m-m'}.\ee

Thus, we finally obtain:

\be \Delta P_m \approx -\frac{P_m}{N-k} + \frac{1}{N-k}\sum_{m'=1..(m-1)} P_{m'} P_{m-m'}, \ee

where the second term of Eq. (\ref {prob}) changed the factor of two appearing in Eq. (\ref{num}) to be one.

It is now straightforward to take the continuous limit, where it is convenient to define a fictitious time related to the step number as $t = - \log \frac{N-k}{N}$. This leads to the following integro-differential equation for the time evolution of the masses:

\be \frac{\partial \mathcal {P}(m,t)}{\partial t} = -\mathcal
{P}(m,t) + \int_{0}^{m} \mathcal {P}(m',t) \mathcal {P}(m-m',t) dm',
\label {partial} \ee where the time $t$ is related to step number
$k$ by the equation $t=-\rm{log}[(N-k)/N]$. This is a particular
case of the Smoluchowski coagulation equation \cite{smoluchowski}.
It can be verified that:

\be \mathcal {P}(m,t)= e^{-e^{-t}m}e^{-t}, \label {RG_solution} \ee
is a solution. We expect this solution to be relevant for $t \gg 1$,
when the slightly different initial conditions it obeys are not
important \cite{aizenman_bak} (initially, all masses are equal, not
exponentially distributed).

\section {4. RG solution in one dimension}
In the following we show how in one dimension one can find $P(\lambda)$ with the RG approach, without using the result of the moment calculation.
Eq. (\ref{RG_solution}) shows that the mass distribution at a given time is given
by $\mathcal {P}(m,t)= e^{-e^{-t}m}e^{-t}$. From this, we can
calculate $\langle 1/m \rangle = \int \frac{\mathcal {P}(m,t)}{m}
dm$. Approximating the exponential as an effective cutoff at
$m_c=e^{t}$, we find that $\langle 1/m \rangle \sim \rm{log}(m_c)
e^{-t} \sim t e^{-t},$ where we took a lower cutoff of order unity
for the mass.

Using the relation between $t$ and $k$ we find that:
 \be \langle 1/m
\rangle \sim (N-k)/N \label {RG_mass} ,\ee where $N-k$ is the number
of masses left after $k$ RG steps.

For the one dimensional case we can combine this result with the
previously obtained renormalization of the springs, to obtain the
mass correction. Since the eigenmodes obey $\omega^2 = K/\mu$, with
$\mu$ the reduced mass of the two current masses, the springs
eliminated at this point of the RG flow, whose distribution was
shown to obey $P(K) \sim 1/K^{1-\epsilon}$, contribute on average to
\be -\lambda = \omega^2 = K\langle 1/m_1+1/m_2 \rangle= 2K \langle
1/m \rangle = 2K^{1+\epsilon}.\ee Calculating the distribution of
$\lambda$ arising from the known spring distribution $P(K)$ gives
\be P(\lambda) \sim 1/ \lambda^\frac{1}{1+\epsilon},
\label{1d_spec}\ee where the lower cutoff is still $\lambda=0$ but
the higher cutoff is $\lambda=-2$. Thus, the effect of the mass RG in one dimension is changing $P(\lambda) \sim 1/\lambda^{1-\epsilon}$
to $1/\lambda^{\frac{1}{1+\epsilon}}$, which is negligible for small
$\epsilon$. For low densities the RG approach and the derivation
via the calculation of the moments give identical results, as they
should.


\begin{thebibliography}{10}
\bibitem{diffusion_laplace}
To see this, one can write the probability to remain in a given site $i$ in
  terms of the eigenmodes and eigenvalues, as: $\hat{P}(t)=\sum_\lambda
  |\langle V_\lambda | i \rangle|^2 e^{-\lambda t}$, with $V_\lambda$ the
  corresponding eigenmodes. Averaging over $i$ gives $\hat{P}(t)=\sum_\lambda
  e^{-\lambda t}$, which, in the continuous limit, is the Laplace transform of
  the probability density.
\end{thebibliography}
\end{document}